# Health literacy in e-oncology care: challenges and strategies

**Authors:**
Hajar Hasannejadasl[1], Cheryl Roumen[1], Yolba Smit[2], Andre Dekker[1], Rianne Fijten[1]

**Affiliations:**
1. Department of Radiation Oncology (Maastro), GROW School for Oncology, Maastricht University Medical Centre+, 6229 ET Maastricht, The Netherlands.
2. Department of Hematology, Radboud ~~U~~university ~~m~~Medical ~~c~~Center, Nijmegen, The Netherlands

**Corresponding author:**
Rianne Fijten
ORCID: https://orcid.org/0000-0002-1964-6317
Email: r.fijten@maastrichtuniversity.nl
Post address: Paul Henri Spaaklaan 1, Maastricht, the Netherlands

**Abstract:**

Given the impact of health literacy (HL) on patients' outcomes, limited health literacy (LHL) is a major barrier in cancer care globally. HL refers to the degree in which an individual is able to acquire, process and comprehend information in a way to be actively involved in their health decisions. Previous research found that almost half of the population in developed countries have difficulties in understanding health related information. With the gradual shift toward the shared decision making (SDM) process and digital transformation in oncology, the need for dealing with low HL issues is more crucial. Decision making in oncology is often accompanied by considerable consequences on patients' lives, which requires patients to understand complex information and be able to compare treatment methods by considering their own values. How health information is perceived by patients is influenced by various factors including patients' characteristics and the way information is presented to patients. Based on the findings, identifying patients with low HL and using simple data visualizations are the best practice to help patients and clinicians in dealing with LHL. Furthermore, preparing reliable sources of information in tools such as patient decision aids (PDA), as well as involving HL mediators in consultation sessions supports patients to make sense of complex information.

## 1. Introduction:

Cancer is the second leading cause of death in the world, with a mortality rate of 10 million in 2020 (1). Given that patients' understanding of their disease and its appropriate treatments is critical to the medical decision making process, HL plays an important role in improving patients' awareness and empowering them (2). HL refers to the set of individual skills to access, process, understand and use health information to stay healthy (3). HL skills are important during a healthy life span, but especially in disease-related decision making. These skills not only aid in preventing disease but also help a person make smart decisions in the event of disease and enable them to manage the disease and associated self-care (4).

Due to the importance of HL in prevention and disease management, LHL has the potential to cause serious problems. Factors associated with HL include the patient's characteristics (5), the communication between physician and patient (6) and the way information is presented (7). As a result, how health-related information is presented has a significant impact on individuals' perceptions of disease risk and, consequently, on how individuals make decisions about their health and related behaviors (4). For example, during the onset of disease, people experience higher levels of stress and receive a lot of new and complex information, which is challenging for patients with LHL. And yet care providers often assume that HL levels are higher than they really are (8). This is further exacerbated due to patients' reluctance to expose their inability to understand health information due to the stigma associated with LHL (8,9). Nowadays, a lot of attention is paid to patient participation in decisions related to their health, and patient centered care. SDM is the process in which caregivers and the patients work together to make the best decision, requires the exchange of information and discussion about treatment options and exploring patients' values (10). However, the inconsistency between the often complex nature of the presented information and the level of HL of the individual influences patient-physician communication negatively and may impact patients' outcomes (9). In addition, how risk is perceived by patients depends on how the information is presented to them (11). LHL is not limited to a certain population or region. In fact, LHL is a common issue around the world (12,13). For instance, almost half of the European population has LHL (14); similar findings have been reported in Australia (15), US and UK (9). Given the impact of socioeconomic status on HL (16), in developing countries HL is remarkably poor (17).

To get the most out of SDM, it is therefore important to consider factors that cause and influence HL and examine strategies to deal with LHL. The goal of this mini-review is to explore the consequences of LHL and propose strategies to effectively deal with patients with LHL during medical decision making in oncology.

## 2. The impact of poor HL skills

A person with good HL is able to identify and access reliable sources of information, as well as differentiate it from misinformation, perceive risk and analyze data about their own situation (19). Good HL is associated with patients' ability to manage their disease in general, their ability to effectively communicate with care providers and participate in the health decision-making process (20,21). Therefore, adequate HL will put the patient on the right care path and eventually, promote the quality of care (22).

In oncology in particular, poor HL could lead to many undesirable outcomes. For instance, wrongfully interpreting risks produced by a clinical decision support system (CDSS) could lead patients and their doctors to choose not the most suitable treatment decision due to misinterpretation of the results (23,24). In turn, wrong decisions could lead to decisional regret later on especially, when there are multiple options that involve trade-offs between harms and benefits (25). In addition, using health services requires some HL skills to compare the benefits and limitations. For example, the aim of screening is to diagnose cancer in the early stage but there is a possibility of harm due to overdiagnosis and false positive or negative results. An example of this is breast and prostate cancer screening, where 92% of women and 89% of men in nine European countries overestimate the benefits of mammography and PSA as markers for breast and prostate cancer respectively (26,27). On the contrary, overestimating risks instead of benefits might lead to unnecessary fear, decreased mental health status (28) and may cause patients to choose the 'safe' (i.e. more invasive but not necessarily more effective) treatment option even though this may increase side effects and negatively influence daily life in a more pronounced way (29).

LHL can also lead to a poorer quality of life (QOL). For instance, people with LHL are more likely to adopt an unhealthy lifestyle such as smoking or lack of exercise which will increase risk of developing chronic diseases (30) and subsequently reduce their QOL (31). Patients with lower HL are also less likely to use preventive health services, resulting in later diagnosis of cancer. This delay in diagnosis and treatment increases the risk of hospitalization (32), reduces the odds of survival and increases the cost of care (31). Due to this cascade of accumulating negative impacts patients with LHL are 1.5 to 3 times more likely to encounter undesirable health events such as higher risk of hospitalization (5). Additionally, certain characteristics make people more susceptible to LHL and influence risk perception. These risk factors include older age, lower educational level, lower socioeconomic status, immigration, having chronic diseases, and physical or mental disabilities (5,8,33).

New technologies such as artificial intelligence (AI) offered many opportunities in healthcare (34). AI methods have the potential to transform medical decision making by learning from vast amounts of data, updating state-of-the-art medical information, increasing the speed of diagnosis and reducing medical errors. In addition, AI is able to predict outcomes related to disease and treatment outcomes (35). Despite its potential, risks presented by AI may be difficult to interpret, especially for people with LHL. Even highly educated physicians trained in dealing with risky decisions have trouble assessing risks (36). In some cases the information presented is too vague or skewed to highlight the benefits, possibly resulting in misunderstanding by its users. For example, deep learning is a powerful form of AI which may outperform physicians in certain situations, but the logic of the method is unclear and may lead to misinformed decision making (37).

### 3. Practical solutions to deal with LHL/ insights for intervention

Due to the severe and cascading effects that LHL can have, strategies are warranted that allow better participation of LHL patients in medical decision making. Here we describe four strategies that could be applied in clinical practice immediately. The first and most important strategy is to identify patients with LHL in order to adapt communication and presentation of health information. The second, third and fourth strategy involve support measures to aid communication and visualization.

Strategy 1: Identifying patients with LHL to tailor SDM accordingly

Since physicians tend to overestimate a patient's HL level, strategies are needed to objectively assess this before they convey health information and risks. A large body of literature concerning this issue and several questionnaires are available for this purpose.

The S-TOFHLA ("Short version of Test Of Functional Health Literacy in Adults") is a questionnaire that assesses the HL level of people specifically in healthcare settings. It consists of two sections for reading and numeracy skills, is available in multiple languages, and can be completed in less than 12 minutes (38). Alternatives for the S-TOFHLA specific to the field of oncology are the Cancer Health Literacy Test (CHLT-30) and CHLT-6 questionnaires which are externally validated and have been found reliable (22).

Despite availability of HL screening instruments, there are many concerns such as lack of time for clinicians and risk of stigma for patients (39). Patients are reluctant to test for HL because of the risk of labelling as illiterate and shame around it. The disadvantages of testing HL may outweigh the benefits as it may keep patients from seeking healthcare services (39). To prevent the risk of demotivating LHL patients, providing a shame free environment is essential. Cornet et al,. (40) proposed an indirect framework for assessing HL by observing patients' behavior and asking questions about their understanding from a piece of text. Inability to give a complete overview of their own medical history and feeling nervous with complex information are some examples of common observed behaviours in individuals with LHL. Recently, in the Netherlands first interactive digital tools have been developed to identify and practice communication to people with LHL (41). Given the fact that no consensus has been reached on the best method to assess HL and given the disadvantages of HL questionnaires, we propose a less formal and multidesign approach. This multidesign approach could consist of a combined e-learning and training to medical professionals to quickly identify patients with LHL. In addition, we propose to identify patients with LHL by using a short version of the HL screening tools and frame it as a way to personalize information. Ideally, the outcome of the test would result in a digital signal to the electronic patient dossier that a patient has a LHL. Identifying HL level is the first step to communicate risk information in a personalized way.

Strategy 2: Including health literacy mediators to guide LHL patients
A literacy mediator is a person who uses their literacy skills to help others in different areas. In the past, this mainly related to reading and writing as only a limited number of people had the ability to read and write. These literacy mediators assisted illiterate people in understanding and communication via written text (42,43). Mediators can also specifically help individuals, especially the vulnerable, with LHL in the (medical) decision making process. These mediators who are willing to use their HL skills to support others are called health literacy mediators (HL mediators). Edwards et al. (44) explored the potential of a personal social network as the function of a health mediator ranges from information seeking, self management to SDM depending on the situation. They also examined the possibility of HL support in training classes which provide the opportunity to discuss with other participants too. A successful example of HL mediators contribution showed in the study by Kosa et al. (45) where the high participation rate indicated the effectiveness of HL mediators in encouraging different groups including LHL people in using preventive service. Therefore, hiring a literacy mediator, especially in areas with a high prevalence of LHL, such as low income regions, is an essential step. In a shame free atmosphere patients with LHL are supported by mediators to talk about what they are struggling to understand.

Strategy 3: Adopting risk visualization aids
The way risks, in particular those presented by AI, are presented to patients is crucial, especially for those with LHL. This risk presentation is two-fold: (i) the way a message is framed, either orally or in written text; and (ii) the way the risk is presented visually.
First, the way a message is framed is an important component of conveying information. Message framing could emphasize benefits or losses based on how the message is written or spoken. Gain framing emphasizes the benefits of a decision or choice, for example by emphasizing cancer survival as a result of a treatment decision. On the other hand, loss framing emphasizes the harms of a decision, for example a patient's risk of dying as a result of a decision instead of its opposite, survival. Which form of message framing is most beneficial depends on the situation. When gain framing is used to describe objectively risky treatments, people are more likely to choose that risky treatment than when it's described with loss framing (46,47). In contrast, loss framing resulted in patients choosing less invasive treatments (48).

A second component of risk presentation is the visual representation of that risk, ideally with visual aids. In general, visual aids will convey information better than conventional numerical formats (33), but there are many options to choose from. The most commonly used visual aids are icon arrays (Figure 1A) and bar charts (Figure 1B).

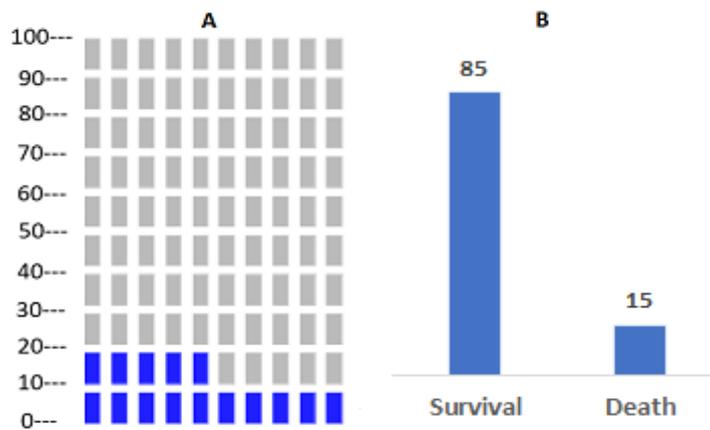

Figure 1. Icon array (A) to bar chart (B)

Icon arrays are a type of visual aid in which a shape is repeated a number of times, generally 100 times, when risks are displayed. A subset of those shapes will be presented differently to indicate a proportion. For example, a risk of 15% would be represented as 15 blue boxes in a field of 100 grey boxes (Figure 1A). This type of visual aid lends itself well to presenting percentages in a larger population.

It is important to note that the way the shapes are presented may influence the users' understanding. Zikmund-Fisher et al. (49) found that anthropomorphic icons, i.e. icons that resemble human or human-like form (Figure 2A), were preferred by most patients and Wangeamsermsuk et al. (50) found that anthropomorphic icons were the preferred format in populations with varying HL levels, but that risk perception in LHL patients was better with shapes that referred to (close to) real-life objects (Figure 2B).

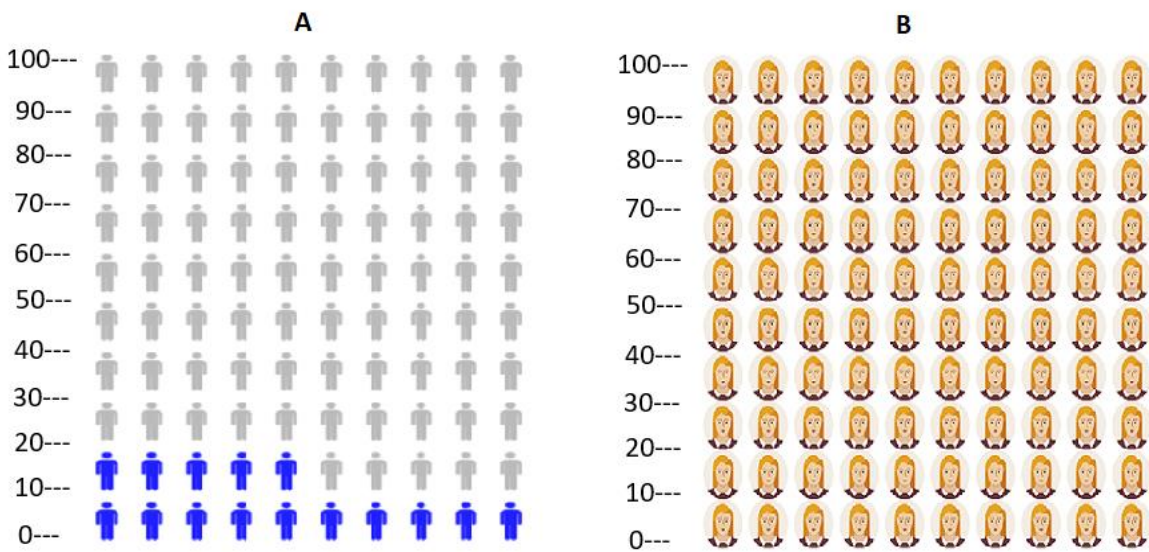

Figure 2. Different types of icon arrays (A) anthropomorphic icons (B) icons representing faces

Bar charts on the other hand are represented as bars of various lengths, where the length of the bar indicates the height of the risk (Figure 1B). This type of visual aid lends itself well to comparing different options, each with associated risks that need to be compared (51). The selection of a type of visual representation of risks depends on the use-case and the intended audience's overall level and variety of HL and demographic characteristics. For example, the eldery (>65 years old) will likely have more difficulty understanding pie charts than those below 65 years of age (52). It will generally also help to use textual information as additional info next to an icon array.

To sum up, due to the high impact of visualization on risk perception, especially patients with low HL, it is important to use well-designed formats such as icon arrays supported by plain language, to convey health information. Additionally, using gain and loss framing at the same time is a practical solution to avoid misinterpretation in this group of patients.

Strategy 4: Using patient information resources
While a wide range of information in different sources and platforms are available for patients, reliability is key for patients to make informed decisions. Medical information sources can take many forms. For example, as a pamphlet prepared by the healthcare institute or the doctor, as an online video developed by a patient advocacy group, or a medical mobile health (mHealth) app on their smartphone, tablet or computer. The goal is to assure that patients fully understand their own risk and empower them for informed decision making. mHealth in particular has become more prominent and is promoting HL skills as well as reducing cost (53,54).

A particular subgroup of information sources relevant to medical decision making is the field of PDAs. These aids are specifically designed to help patients with a particular disease understand relevant information to their disease and possible treatment, and provides them with tools for value clarification (55). These can be provided in a variety of forms as well, including paper and electronic versions (56,57), and have been shown to reduce decisional conflict (58). However, PDAs generally contain large amounts of information, which can be overwhelming for individuals with LHL. In addition, PDAs are not targeted at a population with LHL, but at patients with a general HL. It is therefore important that researchers take special care of making the information presented in a PDA understandable for LHL patients.

As a short-term strategy we propose to provide patients with specific links to reliable sources. To assure the information on this site is reliable, we recommend having a central location with reliable health related information and a quality mark on that website. MedlinePlus for different diseases and kanker.nl specifically for cancer (in Dutch) are examples of reliable information resources.

**Improvement of HL in future healthcare**
The strategies described above provide an overview of the steps healthcare institutes and physicians are able to take immediately to alleviate problems related to LHL of their patients in medical decision making. Other strategies to improve HL of a population in the long term have not been described here, but are nonetheless crucial to medical decision making. Long term strategies include awareness and easy accessibility to terminology lists per health area and health related educational symposia for laymen, educational approaches such as training HL skills in schools for youth, as well as establishing educational materials for other age groups. Raising awareness about the importance of HL in public through for example the media will aid HL in the long term as well. In addition, involving LHL patients in the design and implementation phases of PDA should be a requirement in development of any PDA.

**6. Conclusion**

Despite the emphasis on the importance of HL in the well-being of an individual, LHL is a major issue around the world. LHL may impact all aspects of individuals' life, especially in chronic diseases such as cancer due to the need for identifying symptoms, early diagnosis, and being aware of the probability of side effects, recurrence or survival rate is more essential. We reviewed the current knowledge and have recommended a series of strategies that will help patients, physicians and healthcare organisations better deal with LHL. Future studies are required to assess the utility of provided strategies, as well as explore factors that influence HL skills in the AI era.

**References**
1. Bray F, Ferlay J, Soerjomataram I, Siegel RL, Torre LA, Jemal A. Global cancer statistics 2018: GLOBOCAN estimates of incidence and mortality worldwide for 36 cancers in 185 countries. CA: A Cancer Journal for Clinicians. 2018;68(6):394–424.
2. Improving cancer literacy in Europe to save time, costs and lives (Guest blog) [Internet]. [cited 2021 Sep 2]. Available from: https://www.efpia.eu/news-events/the-efpia-view/blog-articles/improving-cancer-literacy-in-europe-to-save-time-costs-and-lives-guest-blog/
3. Ishikawa H, Kiuchi T. Association of Health Literacy Levels Between Family Members. Front Public Health. 2019 Jun 19;7:169.
4. CDC. The What, Why, and How of Health Literacy [Internet]. Centers for Disease Control and Prevention. 2021 [cited 2021 Sep 17]. Available from: https://www.cdc.gov/healthliteracy/learn/Understanding.html
5. DeWalt DA, Berkman ND, Sheridan S, Lohr KN, Pignone MP. Literacy and Health Outcomes. J Gen Intern Med. 2004 Dec;19(12):1228–39.
6. Street RL. How clinician–patient communication contributes to health improvement: Modeling pathways from talk to outcome. Patient Education and Counseling. 2013 Sep 1;92(3):286–91.
7. Arcia A, Bales ME, Brown W, Co MC, Gilmore M, Lee YJ, et al. Method for the Development of Data Visualizations for Community Members with Varying Levels of Health Literacy. AMIA Annu Symp Proc. 2013 Nov 16;2013:51–60.
8. Local action on health inequalities: improving health literacy [Internet]. GOV.UK. [cited 2021 Sep 16]. Available from: https://www.gov.uk/government/publications/local-action-on-health-inequalities-improving-health-literacy
9. Easton P, Entwistle VA, Williams B. How the stigma of low literacy can impair patient-professional spoken interactions and affect health: insights from a qualitative investigation. BMC Health Serv Res. 2013 Aug 16;13:319.
10. Elwyn G, Frosch D, Thomson R, Joseph-Williams N, Lloyd A, Kinnersley P, et al. Shared Decision Making: A Model for Clinical Practice. J Gen Intern Med. 2012 Oct;27(10):1361–7.
11. Tan SB, Goh C, Thumboo J, Che W, Chowbay B, Cheung YB. Risk perception is affected by modes of risk presentation among Singaporeans. Ann Acad Med Singap. 2005 Mar;34(2):184–7.
12. Nutbeam D. The evolving concept of health literacy. Social Science & Medicine. 2008 Dec 1;67(12):2072–8.

13. Qi S, Hua F, Xu S, Zhou Z, Liu F. Trends of global health literacy research (1995–2020): Analysis of mapping knowledge domains based on citation data mining. PLOS ONE. 2021 Aug 9;16(8):e0254988.
14. Sørensen K, Pelikan JM, Röthlin F, Ganahl K, Slonska Z, Doyle G, et al. Health literacy in Europe: comparative results of the European health literacy survey (HLS-EU). European Journal of Public Health. 2015 Dec 1;25(6):1053–8.
15. Jayasinghe UW, Harris MF, Parker SM, Litt J, van Driel M, Mazza D, et al. The impact of health literacy and life style risk factors on health-related quality of life of Australian patients. Health Qual Life Outcomes. 2016 May 4;14:68.
16. Stormacq C, Van den Broucke S, Wosinski J. Does health literacy mediate the relationship between socioeconomic status and health disparities? Integrative review. Health Promotion International. 2019 Oct 1;34(5):e1–17.
17. Koirala R, Gurung N, Dhakal S, Karki S. Role of cancer literacy in cancer screening behaviour among adults of Kaski district, Nepal. PLOS ONE. 2021 Jul 13;16(7):e0254565.
18. Fernández-Luque L, Bau T. Health and Social Media: Perfect Storm of Information. Healthc Inform Res. 2015 Apr;21(2):67–73.
19. Levin-Zamir D, Bertschi I. Media Health Literacy, eHealth Literacy, and the Role of the Social Environment in Context. Int J Environ Res Public Health. 2018 Aug;15(8):1643.
20. Edwards M, Wood F, Davies M, Edwards A. The development of health literacy in patients with a long-term health condition: the health literacy pathway model. BMC Public Health. 2012 Feb 14;12(1):130.
21. Sørensen K, Makaroff LE, Myers L, Robinson P, Henning GJ, Gunther CE, et al. The call for a strategic framework to improve cancer literacy in Europe. Archives of Public Health. 2020 Jun 23;78(1):60.
22. Dumenci L, Matsuyama R, Riddle DL, Cartwright LA, Perera RA, Chung H, et al. Measurement of Cancer Health Literacy and Identification of Patients with Limited Cancer Health Literacy. J Health Commun. 2014;19(0 2):205–24.
23. Woloshin S, Schwartz LM, Moncur M, Gabriel S, Tosteson ANA. Assessing Values for Health: Numeracy Matters. Med Decis Making. 2001 Oct 1;21(5):382–90.
24. Woloshin S, Schwartz LM, Black WC, Welch HG. Women's Perceptions of Breast Cancer Risk: How You Ask Matters. Med Decis Making. 1999 Aug 1;19(3):221–9.
25. Joyce DD, Heslop DL, Umoh JI, Brown SD, Robles JA, Wallston KA, et al. Examining the association of health literacy and numeracy with prostate-related knowledge and prostate cancer treatment regret. Urologic Oncology: Seminars and Original Investigations. 2020 Aug 1;38(8):682.e11-682.e19.
26. Wegwarth O, Gigerenzer G. The Barrier to Informed Choice in Cancer Screening: Statistical Illiteracy in Physicians and Patients. Recent Results Cancer Res. 2018;210:207–21.
27. Gigerenzer G, Mata J, Frank R. Public Knowledge of Benefits of Breast and Prostate Cancer Screening in Europe. J Natl Cancer Inst. 2009 Sep 2;101(17):1216–20.
28. Hawley ST, Janz NK, Griffith KA, Jagsi R, Friese CR, Kurian AW, et al. Recurrence risk perception and quality of life following treatment of breast cancer. Breast Cancer Res Treat. 2017 Feb;161(3):557–65.
29. Boss EF, Mehta N, Nagarajan N, Links A, Benke JR, Berger Z, et al. Shared decision-making and choice for elective surgical care: A systematic review. Otolaryngol Head Neck Surg. 2016 Mar;154(3):405–20.
30. Froze S, Arif MT, R. S. Determinants of Health Literacy and Healthy Lifestyle against Metabolic Syndrome among Major Ethnic Groups of Sarawak, Malaysia: A Multi-Group Path Analysis. The Open Public Health Journal [Internet]. 2019 Apr 30 [cited 2021 Sep 27];12(1). Available from: https://openpublichealthjournal.com/VOLUME/12/PAGE/172/
31. Akanuwe JNA, Black S, Owen S, Siriwardena AN. Communicating cancer risk in the primary care consultation when using a cancer risk assessment tool: Qualitative study with service users and practitioners. Health Expectations. 2020;23(2):509–18.
32. Samoil D, Kim J, Fox C, Papadakos JK. The importance of health literacy on clinical

cancer outcomes: a scoping review. Ann Cancer Epidemiol. 2021 Mar;5:3–3.
33. Garcia-Retamero R, Cokely ET. Designing Visual Aids That Promote Risk Literacy: A Systematic Review of Health Research and Evidence-Based Design Heuristics. Hum Factors. 2017 Jun 1;59(4):582–627.
34. Davenport T, Kalakota R. The potential for artificial intelligence in healthcare. Future Healthc J. 2019 Jun;6(2):94–8.
35. Jiang F, Jiang Y, Zhi H, Dong Y, Li H, Ma S, et al. Artificial intelligence in healthcare: past, present and future. Stroke Vasc Neurol [Internet]. 2017 Dec 1 [cited 2021 Sep 3];2(4). Available from: https://svn.bmj.com/content/2/4/230
36. Gigerenzer G, Gaissmaier W, Kurz-Milcke E, Schwartz LM, Woloshin S. Helping Doctors and Patients Make Sense of Health Statistics. Psychol Sci Public Interest. 2007 Nov;8(2):53–96.
37. Gulum MA, Trombley CM, Kantardzic M. A Review of Explainable Deep Learning Cancer Detection Models in Medical Imaging. Applied Sciences. 2021 Jan;11(10):4573.
38. Náfrádi L, Papp-Zipernovszky O, Schulz PJ, Csabai M. Measuring functional health literacy in Hungary: Validation of S-TOFHLA and Chew screening questions. Cent Eur J Public Health. 2019 Dec;27(4):320–5.
39. Kronzer VL. Screening for health literacy is not the answer. BMJ. 2016 Jul 5;354:i3699.
40. Assessing and Addressing Health Literacy - ProQuest [Internet]. [cited 2021 Nov 26]. Available from: https://www.proquest.com/openview/2c40de9769450c2774f1b5e6b5a39217/1?pq-origsite=gscholar&cbl=43860
41. Modules - Goed Begrepen [Internet]. [cited 2022 Jan 6]. Available from: https://goedbegrepen.dialoguetrainer.com/
42. Papen U. Literacy mediators, scribes or brokers? Langage et societe. 2010 Sep 20;n° 133(3):63–82.
43. Papen U. Literacy, Learning and Health – A social practices view of health literacy. LNS. 2009 Oct 9;19–34.
44. Edwards M, Wood F, Davies M, Edwards A. 'Distributed health literacy': longitudinal qualitative analysis of the roles of health literacy mediators and social networks of people living with a long-term health condition. Health Expectations. 2015;18(5):1180–93.
45. Kósa K, Katona C, Papp M, Fürjes G, Sándor J, Bíró K, et al. Health mediators as members of multidisciplinary group practice: lessons learned from a primary health care model programme in Hungary. BMC Family Practice. 2020 Jan 28;21(1):19.
46. Armstrong K, Schwartz JS, Fitzgerald G, Putt M, Ubel PA. Effect of Framing as Gain versus Loss on Understanding and Hypothetical Treatment Choices: Survival and Mortality Curves. Med Decis Making. 2002 Feb 1;22(1):76–83.
47. Moxey A, O'Connell D, McGettigan P, Henry D. Describing treatment effects to patients. J GEN INTERN MED. 2003 Nov 1;18(11):948–59.
48. Almashat S, Ayotte B, Edelstein B, Margrett J. Framing effect debiasing in medical decision making. Patient Educ Couns. 2008 Apr;71(1):102–7.
49. Zikmund-Fisher BJ, Witteman HO, Dickson M, Fuhrel-Forbis A, Kahn VC, Exe NL, et al. Blocks, Ovals, or People? Icon Type Affects Risk Perceptions and Recall of Pictographs. Med Decis Making. 2014 May;34(4):443–53.
50. Department of Industrial Engineering, Chulalongkorn University, Bangkok, Thailand, Wangeamsermsuk S, Jiamsanguanwong A. The Comparison of Iconicity Level of Icon Arrays on Risk Perception. JAIT. 2018;9(4):97–101.
51. Waters EA, Maki J, Liu Y, Ackermann N, Carter CR, Dart H, et al. Risk ladder, table, or bulleted list? Identifying formats that effectively communicate personalized risk and risk reduction information for multiple diseases. Med Decis Making. 2021 Jan;41(1):74–88.
52. Preference for and understanding of graphs presenting health risk information. The role of age, health literacy, numeracy and graph literacy - ScienceDirect [Internet]. [cited 2021 Nov 1]. Available from: https://www.sciencedirect.com/science/article/pii/S0738399120303499

53. Dunn P, Hazzard E. Technology approaches to digital health literacy. International Journal of Cardiology. 2019 Oct 15;293:294–6.
54. Levin-Zamir D, Peterburg Y. Health literacy in health systems: perspectives on patient self-management in Israel. Health Promotion International. 2001 Mar 1;16(1):87–94.
55. Gheondea-Eladi A. Patient decision aids: a content analysis based on a decision tree structure. BMC Medical Informatics and Decision Making. 2019 Jul 19;19(1):137.
56. Centre (UK) NM and P. Patient decision aids used in consultations involving medicines [Internet]. Medicines Optimisation: The Safe and Effective Use of Medicines to Enable the Best Possible Outcomes. National Institute for Health and Care Excellence (UK); 2015 [cited 2021 Oct 31]. Available from: https://www.ncbi.nlm.nih.gov/books/NBK355917/
57. Baptista S, Sampaio ET, Heleno B, Azevedo LF, Martins C. Web-Based Versus Usual Care and Other Formats of Decision Aids to Support Prostate Cancer Screening Decisions: Systematic Review and Meta-Analysis. Journal of Medical Internet Research. 2018 Jun 26;20(6):e9070.
58. Ankolekar A, Vanneste BGL, Bloemen-van Gurp E, van Roermund JG, van Limbergen EJ, van de Beek K, et al. Development and validation of a patient decision aid for prostate Cancer therapy: from paternalistic towards participative shared decision making. BMC Med Inform Decis Mak. 2019 Jul 11;19(1):130.